\documentclass[final,3p,times]{elsarticle}

\usepackage{amssymb}

\usepackage{amsmath}
\usepackage{algorithm}
\usepackage[noend]{algpseudocode}
\usepackage{multirow}

\usepackage{color-edits}
\addauthor{chus}{red}
\addauthor{dmt}{blue}

\journal{Elsevier's Journal}

\begin{document}

\begin{frontmatter}



\title{Ontology Matching with Large Language Models and Prioritized Depth-First Search} 


\author[usc1]{Maria Taboada\corref{cor1}} 
\ead{maria.taboada@usc.es}
\cortext[cor1]{Corresponding author}
\author[usc2]{Diego Martinez} 
\author[usc1]{Mohammed Arideh} 
\author[usc3]{Rosa Mosquera} 
\affiliation[usc1]{organization={Department of Electronics and Computer Science, University of Santiago de Compostela},
             city={Santiago de Compostela},
             postcode={15701},
            country={Spain}
}

\affiliation[usc2]{organization={Department of Applied Physics, University of Santiago de Compostela},
            addressline={}, 
             city={Santiago de Compostela},
             postcode={15701},
            country={Spain}}
\affiliation[usc3]{organization={Department of Functional Biology, University of Santiago de Compostela},
            addressline={}, 
             city={Santiago de Compostela},
             postcode={15701},
            country={Spain}}

\begin{abstract}

Ontology matching (OM) plays a key role in enabling data interoperability and knowledge sharing. Recently, methods based on Large Language Model (LLMs) have shown great promise in OM, particularly through the use of a \textit{retrieve-then-prompt} pipeline. In this approach, relevant target entities are first retrieved and then used to prompt the LLM to predict the final matches. Despite their potential, these systems still present limited performance and high computational overhead. To address these issues, we introduce MILA, a novel approach that embeds a \textit{retrieve-identify-prompt} pipeline within a prioritized depth-first search (PDFS) strategy. This approach efficiently identifies a large number of semantic correspondences with high accuracy, limiting LLM requests to only the most borderline cases. We evaluated MILA using three challenges from the 2024 edition of the Ontology Alignment Evaluation Initiative. Our method achieved the highest F-Measure in five of seven unsupervised tasks, outperforming state-of-the-art OM systems by up to 17$\%$. It also performed better than or comparable to the leading supervised OM systems. MILA further exhibited task-agnostic performance, remaining stable across all tasks and settings, while significantly reducing runtime. These findings highlight that high-performance LLM-based OM can be achieved through a combination of programmed (PDFS), learned (embedding vectors), and prompting-based heuristics, without the need of domain-specific heuristics or fine-tuning.
\end{abstract}


\begin{highlights}
\item Propose a novel retrieve-identify-prompt pipeline for ontology matching.
\item Achieve the highest F-Measure score in five of the seven tasks in the unsupervised setting.
\item Exhibit task-agnostic high performance, remaining stable across all tasks and settings.
\item Significantly reduce the number of requests to the LLM.
\end{highlights}


\end{frontmatter}



\section{Introduction}\label{intro}
In the field of information management, ontologies play a key role in semantic interoperability and knowledge exchange by providing a shared vocabulary that promotes common understanding within a domain. They act as valuable resources in various AI applications, such as autonomous communication in manufacturing environments \citep{sapel2024review}, automation of information flows \citep{esteves2024analysis}, smart contract creation \citep{dominguez2024role}, or 3D scene graph generation \citep{strader2024indoor}. However, the proliferation of incomplete and overlapping ontologies within the same domain is an obstacle to smooth communication between applications. In this context, Ontology Matching (OM) becomes essential for integrating distributed knowledge \citep{euzenat2007ontology}. OM has broad applications, from linking entities across public ontologies to facilitating knowledge integration in collaborative business environments \citep{osman2021ontology}, and merging disparate data warehouses for transactional or analytical purposes in private corporations \citep{portisch2022background}. Furthermore, OM shares many similarities with knowledge graph alignment \citep{zhao2020experimental}, and database schema matching \citep{zhang2023schema}. Therefore, advances in OM can be leveraged in a wide range of fields beyond ontologies.

OM refers to the process of identifying semantic correspondences between entities across multiple ontologies \citep{euzenat2007ontology}. Since 2004, the Ontology Alignment Evaluation Initiative (OAEI) has been organizing annual evaluation campaigns to evaluate and benchmark OM technologies, significantly advancing the field \citep{jimenez2025proceedings}. These evaluation campaigns have been instrumental in increasing the performance of existing approaches, particularly those focused on identifying equivalence correspondences between entities of a pair of ontologies - referred to as simple pairwise equivalence OM \citep{osman2021ontology}. Consequently, current OM systems have achieved performance improvements, with techniques
based on lexical, structural and semantic matching \citep{chen2024ontology}, as well as mapping repair techniques \citep{jimenez2011logmap,jimenez2017logmap}. Despite these advances, OM systems continue to have difficulty distinguishing between entities that are semantically similar and those that are merely frequently co-occurring \citep{kolyvakis2018biomedical,he2022machine}. Moreover, the scalability problem prevents widespread implementation \citep{chen2024ontology}. 

\subsection{Motivation and main contributions}\label{motivation}
Recent OM systems often require fine-tuning with large domain-specific training datasets to achieve optimal performance \citep{chen2021augmenting,he2022bertmap,faria2023results,gosselin2023sorbet}. In contrast, Large Language Model (LLM)-based methods have emerged as a promising alternative for OM. They leverage pre-trained knowledge for finding correspondences across ontologies and do not require fine-tuning. However, querying LLMs for all entity pairs results in a quadratic time complexity of $O(n^2)$, where $n$ is the number of entities, making this approach impractical for large datasets. To mitigate this issue, state-of-the-art LLM-based systems apply a \textit{retrieve-then-prompt} pipeline. In this method, $k$ relevant target entities are first retrieved and then used to prompt an LLM to predict mapping correspondences \citep{hertling2023olala,peeters2023using,wang2023exploring, giglou2024llms4om}. This approach reduces the time complexity to $O(n \cdot k)$, significantly improving scalability \citep{giglou2024llms4om}. Despite these advancements, these systems still face challenges, particularly in complex tracks where performance in terms of F-Measure is limited. Furthermore, while LLM-based methods are technically feasible, they face scalability issues due to the long execution times required to query the LLM, particularly when applied to large-scale ontologies in real-world scenarios and when using resource-intensive LLMs \citep{hertling2023olala}.

To address these challenges, we propose MILA (\underline{MI}nimizing \underline{L}LM Prompts in Ontology M\underline{A}pping), a framework designed to improve F-Measure performance and eliminate unnecessary queries to LLMs. MILA introduces a novel \textit{retrieve-identify-prompt} pipeline, which adds an intermediate step to identify high-confidence bidirectional (HCB) correspondences with high precision. This step involves simple heuristics that operate in constant time, $O(1)$, eliminating $k$ LLM queries for each identified HCB correspondence, then reducing LLM interactions to only borderline cases. For these edge correspondences, MILA applies a prioritized depth-first search (PDFS) strategy, which iteratively queries the LLM until a definitive match is confirmed. This approach also minimizes the number of queries by ending early when a valid match is found. Although the overall time complexity remains comparable to existing LLM-based OM systems, MILA significantly reduces execution time, especially when the retrieval system ranks the most relevant candidates first. To achieve efficiency gains, MILA's retrieval system leverages the SBERT embedding model \citep{reimers2019sentence}, prioritizing correspondences between entities that have the most semantically similar names.

Our framework was evaluated using the biomedical challenge proposed in the 2024 edition of the Ontology Alignment Evaluation Initiative (OAEI) \citep{jimenez2025proceedings}. We selected the biomedical domain due to its particular nature, which makes OM a challenge \citep{kolyvakis2018biomedical}: the presence of a rich and constantly evolving terminology, the significant variability in language usage, and the high frequency of rare terms, which are difficult to learn. Moreover, the limited performance of LLM-based solutions in this domain highlights the need for new approaches \citep{giglou2024llms4om}. To further validate MILA's robustness and its applicability beyond the biomedical domain, we also assessed its performance on two tasks from the anatomy and biodiversity challenges from the 2024 edition. The results of our evaluation demonstrate the ability of MILA to outperform state-of-the-art OM systems in terms of task-agnostic and high performance in terms of F-measure. They also demonstrate reduced execution times compared to state-of-the-art LLM-based OM systems. These excellent results corroborate the strength of our proposal. 

\section{Related work}\label{relatedwork}
OM technology requires the use of external background knowledge to work effectively, as most ontologies are designed in specific contexts that are not explicitly modeled \citep{portisch2022background}. The most recent OM systems consume existing pre-trained neuronal models \citep{vaswani2017attention} as sources of external knowledge \citep{chen2021augmenting,he2022bertmap,faria2023results,gosselin2023sorbet}. These systems show significant performance improvements over traditional feature engineering and machine learning strategies. They benefit from the capabilities that pre-trained neural models have to automatically interpret the textual descriptions embedded in the labels, comments and definitions of ontologies. However, most models need to be fine-tuned with large training data to perform properly, and they can only process short textual descriptions \citep{hertling2023olala}. 

To overcome the aforementioned drawbacks, several studies have explored the promising capabilities of LLMs for OM \citep{hertling2023olala,peeters2023using,wang2023exploring,giglou2024llms4om,norouzi2023conversational,he2023exploring}. All of these studies focus on comparing the effect of different prompt inputs to LLMs \citep{liu2023pre} on OM. In \citep{norouzi2023conversational}, the LLM is provided with complete ontologies in a single prompt, along with detailed instructions on the problem definition and the query goal \citep{norouzi2023conversational}. However, the most commonly used strategy is to include only a pair of ontology entities in each individual prompt \citep{hertling2023olala,peeters2023using,wang2023exploring,giglou2024llms4om,he2023exploring}. Moreover, the performance of LLMs has been studied in both zero-shot and few-shot settings \citep{peeters2023using}. In zero-shot scenarios, where LLMs are queried without providing in-context examples, the performance increases when the prompt contains a set of explicit matching rules. Surprisingly, this zero-shot scenario is almost as effective as providing examples that are textually close to the entities to be matched (few-shot setting) \citep{wang2023exploring}. In addition, prompting LLMs with a brief description of the OM task followed by positive and negative examples has achieved the best results in the anatomy domain \citep{hertling2023olala}. Moreover, the use of multiple choice prompts notably reduces the query execution time, but degrades results. Furthermore, the inclusion of structural context in LLM prompting does not improve OM \citep{he2023exploring}. 

To meet the challenge of large-scale OM \citep{jimenez2011logmap,khoudja2022deep}, LLM-based OM systems integrate Retrieval-Augmented Generation (RAG) \citep{gao2023retrieval} to effectively reduce the problem of incorrect content generation \citep{hertling2023olala,wang2023exploring,giglou2024llms4om}. They follow a naive methodology based on a \textit{retrieve-then-prompt} pipeline that includes three sequential steps \citep{gao2023retrieval}: indexing the target ontology in a vector database, retrieving relevant target entities, and prompting the LLM with mapping candidates. Although the results show that the integration of RAG with LLM in OM is in some cases comparable or even better than current OM systems, the F-Measure performance of LLM-based approaches still requires improvement \citep{ hertling2023olala,wang2023exploring,giglou2024llms4om}. Although these systems have a high candidate recovery rate, which can reach 100\(\%\), the results can decrease by up to 30\(\%\) after prompting. Even this reduction can reach 50\(\%\) in the biomedical domain, where LLM-based systems perform weakly \citep{giglou2024llms4om}. Moreover, although RAG-based approaches significantly reduce time complexity from quadratic to linear  \citep{giglou2024llms4om}, they still require a long runtime when applied to large ontologies \citep{hertling2023olala}. This underscores the need for alternative RAG strategies that improve F-Measure performance while minimizing the number of requests to the LLM \citep{giglou2024llms4om}. One suggested solution is to initially use a fast and highly accurate matcher to find correspondences for straightforward cases, reserving LLM prompting only for more complex or ambiguous cases \citep{hertling2023olala}. This proposal would optimize efficiency by reducing the dependence on the LLM for simple matching tasks. Therefore, the challenge now is to design the matcher so that it can effectively and reliably identify correspondences with minimal computational overhead.

\section{Preliminaries}\label{preli}
\subsection{Problem formulation}
Ontology matching (OM) is the process of identifying semantic correspondences among entities of overlapping ontologies \citep{osman2021ontology}. A simple pairwise OM can be defined by a function that takes as input a source ontology \textit{O\textsubscript{S}}, a target ontology \textit{O\textsubscript{T}}, an input alignment \textit{A}, a set of parameters \textit{p} and resources \textit{r} (such as external background knowledge), and return an alignment \textit{A'} (i.e., a set of correspondences) between entities (or classes) of the input ontologies \citep{euzenat2007ontology}. A semantic correspondence is a quintuple \textit{$<$id, e\textsubscript{O\textsubscript{s}}, r, e\textsubscript{O\textsubscript{T}}, c$>$}, such as:
\begin{itemize}
\item \textit{id} identifies the correspondence,
\item \textit{e\textsubscript{O\textsubscript{S}}} and \textit{e\textsubscript{O\textsubscript{T}}} are entities of \textit{O\textsubscript{S}} and \textit{O\textsubscript{T}}, respectively,
\item \textit{r} identifies the semantic relation between \textit{e\textsubscript{O\textsubscript{S}}} and \textit{e\textsubscript{O\textsubscript{T}}},
\item \textit{c} is a confidence value that reflects the degree of correctness of the correspondence, which is usually a real value in the interval [0,1]. 
\end{itemize}

In our approach, \textit{r} is an equivalence relation (\( \equiv \)) that links two entities through a bidirectional correspondence. Therefore, \textit{$<$id\textsubscript{i},e\textsubscript{O\textsubscript{S}},r,e\textsubscript{O\textsubscript{T}},c$>$} is the inverse of \textit{$<$id\textsubscript{j},e\textsubscript{O\textsubscript{T}},r,e\textsubscript{O\textsubscript{S}},c$>$} \citep{osman2021ontology}. An example of a simple and bidirectional pairwise correspondence between the ontologies NCI Thesaurus (NCIT) \citep{coronado2004nci} and Disease Ontology (DOID) \citep{schriml2012disease}, expressed in Description Logic (DL), is the following:

\begin{center}
\textit{O\textsubscript{NCIT} : clear cell sarcoma of soft tissue  \( \equiv \) O\textsubscript{DOID} : clear cell sarcoma}.
\end{center}

\subsection{Prioritized Depth-First Search (PDFS)}\label{PDFS}

State-space search algorithms aim to find solutions to problems represented by a set of states. They organize the search space into a tree and evaluate the best path based on certain criteria, typically optimizing the cost to reach a goal. Search algorithms are divided into uninformed and informed categories. Uninformed algorithms, such as depth-first search (DFS), explore the state space without knowing how promising a state is. Informed algorithms, such as the greedy best-first search (GBFS), use a heuristic function to guide the search toward the goal by prioritizing nodes that appear closest to the goal. The term \textit{greedy} usually denotes that the decision is never revised. However, sometimes this term is also used to describe an algorithm that backtracks when a dead end is reached, combining elements of both DFS and GBFS. Sometimes, this combined strategy is called \textit{prioritized depth first} search (PDFS) to clearly distinguish it from a pure greedy approach. 

\subsection{Main OM components in RAG-based approaches}

In RAG systems, domain knowledge is stored in vector databases, which are specifically designed to store and index individual text units based on their corresponding embedding vectors. These embedding representations enable the retrieval of relevant information when a query encoded in the same latent space is processed. The retrieval process is usually supported by the semantic similarity between the query and the indexed vectors, facilitating the extraction of contextually relevant text units to query the LLM.

The workflow of most RAG-based OM systems typically involves several sequential steps \citep{hertling2023olala,wang2023exploring,giglou2024llms4om}: 1) \textit{Vector knowledge base (KB) construction}, where the target ontology is extracted, pre-processed, and indexed; 2) \textit{Mapping prediction}, where mapping candidates are retrieved by computing the semantic similarity between the vector representation of a source entity and the indexed target entities; 3) \textit{Mapping refinement}, where candidates are either confirmed or discarded through prompting an LLM; 4) \textit{Candidate filtering}, where mappings are filtered out based on predefined heuristics.

\subsubsection{Vector KB construction}
Based on the assumption that the matched entities are likely to have labels with overlapping subtokens, traditional methods implement the inverted word-level index \citep{jimenez2011logmap,he2022bertmap}. In these approaches, the initial set of correspondences for a source entity is built by selecting target labels that share at least one sub-word token with some label of the source entity. Unlike these approaches, RAG-based OM methods encode and index complete labels and definitions, rather than their sub-words \citep{hertling2023olala}. Label indexing aims at an efficient approach to retrieval, whereas definition indexing is intended to retrieve entities that are not similar in appearance \citep{wang2023exploring}. Hierarchical contexts (parent and child labels) can also be extracted and encoded, although they show worse performance \citep{giglou2024llms4om}. Therefore,  in line with previous works \citep{hertling2023olala,giglou2024llms4om}, our approach only indexes labels (preferred terms and synonyms), with the goal of maximizing retrieval efficiency. 

Moreover, some OM approaches index only the target ontology \citep{wang2023exploring,giglou2024llms4om}, while others index both ontologies, aiming to increase the initial set of candidate correspondences and thus the recall of the approach \citep{he2022bertmap,hertling2023olala}. MILA also indexes both ontologies, but with the goal of properly handling all the search space across them.

\subsubsection{Mapping prediction}
An embedding-based retrieval model predicts the most similar candidates for the correspondences in OM. By computing the cosine similarity between a vector representing the source entity and each vector in the target database, the most similar alignment candidates are predicted. From these, the top k candidates per entity are selected \citep{hertling2023olala,giglou2024llms4om}. 

\subsubsection{Mapping refinement}
In RAG-based approaches \citep{hertling2023olala,wang2023exploring,giglou2024llms4om}, an LLM filters the candidates proposed by the retriever. Most approaches verbalize each alignment candidate in text, which is used to populate an LLM prompt template, following some prompting technique \citep{liu2023pre}. Depending on the type of information provided to the LLM, these techniques can be classified as zero-shot or few-shot. In zero-shot scenarios, LLMs are prompted with no contextual examples provided \citep{hertling2023olala,giglou2024llms4om,he2023exploring}, while a few contextual examples are provided in few-shot settings \citep{hertling2023olala,wang2023exploring}. In the latter case, it may also be appropriate to provide both examples of correct correspondences and missing correspondences \citep{hertling2023olala}. 
  
Based on the amount and type of information provided, prompts can be categorized into two types: simple prompts and contextual prompts. Simple prompts provide minimal ontology context, including only entity names \citep{hertling2023olala}. In contrast, contextual prompts include additional and relevant information, such as definitions or hierarchical relationships \citep{giglou2024llms4om,he2023exploring}. This contextual information can be retrieved directly from the ontologies themselves or from external resources. Alternatively, contextual data can be selectively extracted using graph search algorithms that identify and focus on the most relevant ontology information, leading to more accurate and effective LLM answers \citep{sampels2024exploring}. Furthermore, prompt templates can involve binary decisions, where the LLM must decide whether a candidate is correct or not \citep{hertling2023olala,giglou2024llms4om,he2023exploring}, or multiple decisions, where the LLM must select among several candidate proposals \citep{hertling2023olala,wang2023exploring}.

In approaches prior to RAG \citep{jimenez2011logmap,he2022bertmap,jimenez2020dividing}, mapping refinement aims to discover new correspondences from predicted mappings. These approaches are based on the locality principle \citep{jimenez2011logmap}, which assumes that entities semantically related to those in a predicted correspondence are likely to be matched in new mappings. LogMap \citep{jimenez2011logmap} computes new mappings by expanding the hierarchical contexts of each entity in a mapping and finding correspondences between classes of these two hierarchical contexts. BERTMap \citep{he2022bertmap} also improves the performance of a BERT classifier with a reasonable time cost by restricting this principle to correspondences that have been predicted with a high score.

\subsubsection{Candidate filtering}

Mapping refinement is usually finished with a post-processing step mainly aimed at filtering out incorrect correspondences. OM systems often use methods based on heuristics, such as confidence score thresholds (confidence filters) or criteria to achieve unambiguous alignments (cardinality filters) \citep{faria2023results,hertling2023olala,giglou2024llms4om}. More sophisticated methods rely on logical reasoning \citep{jimenez2011logmap} to remove correspondences that cause logical inconsistencies after integrating ontologies \citep{he2022bertmap,jimenez2013evaluating,santos2015ontology}, or on probabilistic reasoning to resolve conflicts \citep{li2020repairing}. However, our approach does not apply any post-processing step.

\section{Methodological framework}\label{metho}
\begin{figure} [ht]
\centering
\includegraphics[width=\linewidth]{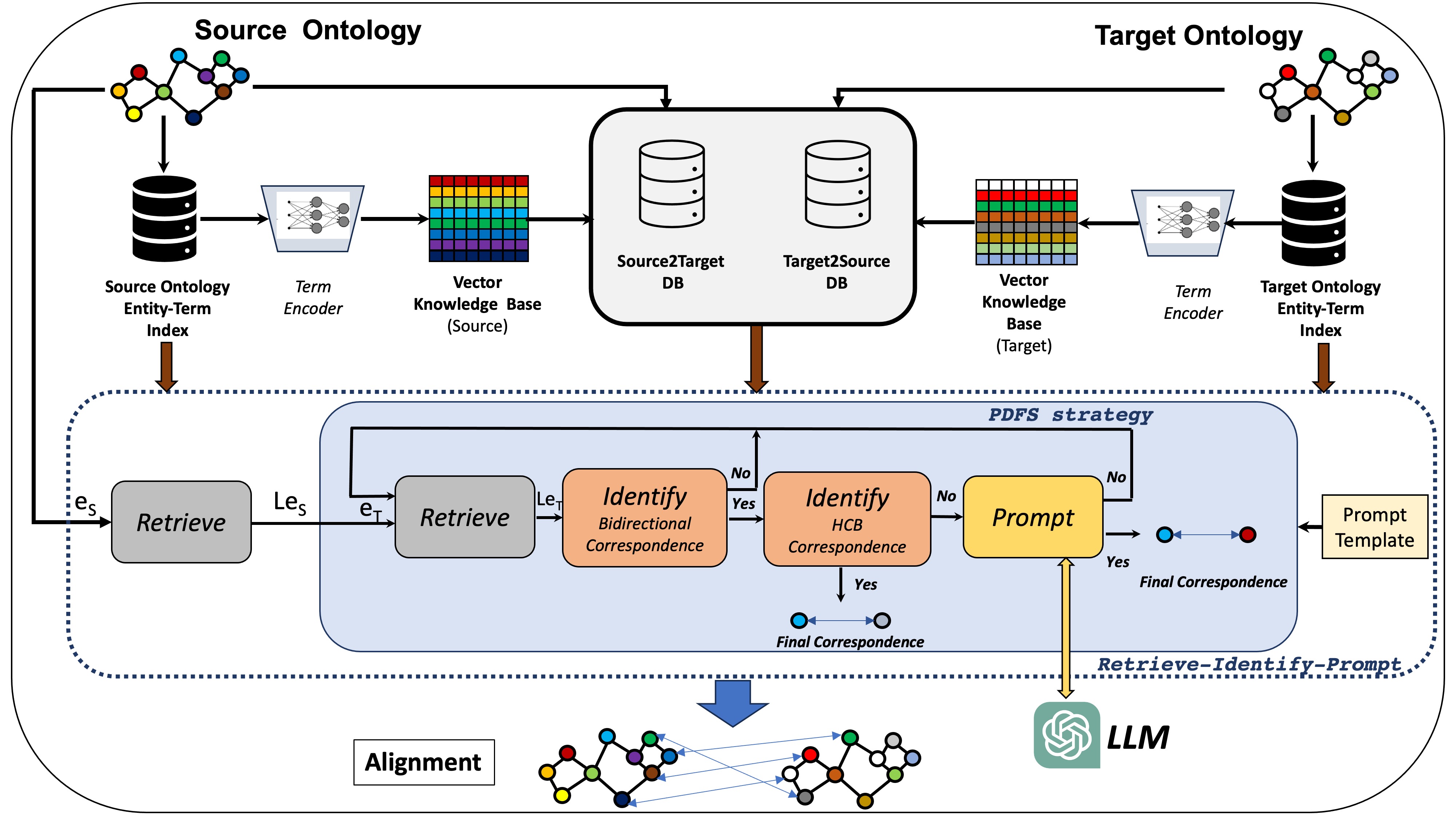}
\caption{Overview of MILA.}\label{Fig:overview}
\end{figure}

Our approach MILA aims to find simple and bidirectional pairwise correspondences. To achieve this, MILA solves the OM task through a novel \textit{retrieve-identify-prompt} pipeline, enhanced by a PDFS strategy. The following subsections outline the key components of our approach and illustrate them with examples. An overview of the main steps is presented in Fig. \ref{Fig:overview}.

\subsection{Vector KB construction}
Given an ontology $O$, an entity (or class) $e \in O$ can have multiple labels (terms) defined by annotation properties, typically including preferred and alternative labels (synonyms). MILA parses the input ontologies and indexes their entities in the \textit{Entity-Term Index} (see Fig. \ref{Fig:overview}). This index comprises three tables: entity-term relations, preferred term-entity relations, and term-entity relations. This structured indexing enables fast access and efficient searching for MILA’s other components.

Let $\Omega(e)$ be the set of labels of an entity $e$. Using an embedding model, such as SBERT \citep{reimers2019sentence}, the encoder maps each label $\omega \in \Omega(e)$ to a vector representation, denoted by $v(\omega) \in {R}^d$, where $v(\omega)$ is the vector representation of the label $\omega$ in a $d$-dimensional vector space (see Fig. 1, \textit{Term Encoders}). Following prior work \citep{hertling2023olala,wang2023exploring,giglou2024llms4om}, we encode full labels to generate the vector KBs. However, we do not incorporate ontology contexts in these KBs. This decision is motivated by the fact that test datasets do not provide definitions for most ontologies \citep{jimenez2025proceedings}, and that using only the ontology’s terminology is an efficient retrieval strategy in terminology-intensive domains, such as the biomedical field \citep{wang2023exploring}. 

\subsection{Mapping prediction}\label{retrieve}
For each source and target entity, MILA pre-computes the most promising correspondence candidates, taking into account the vector representations of their labels. Let $O_S$ and $O_T$ represent source and target ontologies, respectively. Let $e_{S}$ and $e_{T}$ denote source and target entities, respectively. For a given source label $\omega_{S} \in \Omega(e_{S})$ and a target label $\omega_{T} \in \Omega(e_{T})$, the function $f_{w}$ computes the similarity score $sim$ between their respective vector representations, $v(\omega_{S})$  and $v(\omega_{T})$, using a similarity metric. Specifically, the function $f_{w}$ is defined as
\begin{align*}
f_{w}(\omega_{S}, \omega_{T}) &= sim(v(\omega_{S}), v(\omega_{T})).
\end{align*}

\begin{figure}[ht]
\centering
\includegraphics[width=\linewidth]{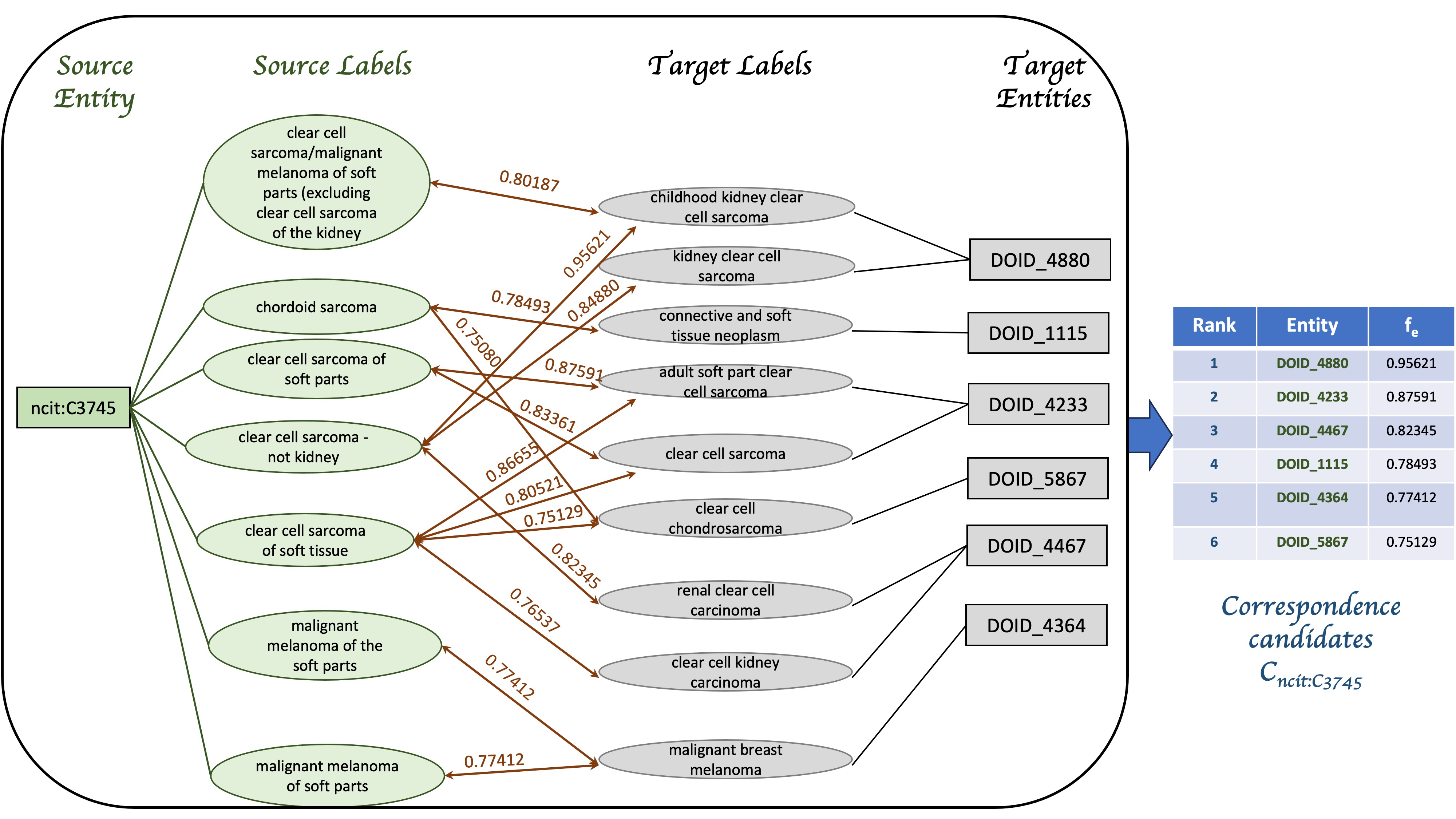}
\caption{Prediction of correspondence candidates for the entity ncit:C3745 (\textit{clear cell sarcoma of soft tissue}) using SBERT.}\label{Fig:Candidateexample}
\end{figure}

We used cosine similarity in this step of the RAG model since it guarantees the retrieval of contextually relevant candidates, even when their labels differ, which is essential to achieve high-quality correspondences \citep{lewis2020retrieval}. Fig. \ref{Fig:Candidateexample} shows the predicted correspondence candidates for the source entity ncit:C3745 (Clear Cell Sarcoma of Soft Tissue) using SBERT. In this figure, the similarity score $f_{w}$ is depicted in brown, representing the similarity between source labels (in green) and target labels (in gray). Specifically, for the terms $\omega_{NCIT}=$ \textit{clear cell sarcoma of soft tissue} and $\omega_{DOID}=$ \textit{clear cell sarcoma}, $f_{w}$ is
\begin{align*}
f_{w}(\omega_{NCIT},\omega_{DOID})=0.80521.
\end{align*}

Let $\Omega(O_T)$ be the set of all labels in $O_T$. Given $k$, for each source label $\omega_{S} \in \Omega(e_{S})$, MILA generates the $k$ most promising correspondence candidates by selecting a subset $C_{\omega_{S}} =\{\omega_{T_1},\omega_{T_2},...,\omega_{T_k} \}  \subseteq \Omega(O_{T})$ that maximizes the score function $f_w$ with respect to $\omega_S$. To refine this selection, we introduce a threshold $\tau$ that filters out candidates whose score function values fall below it. Therefore, the similarity scores satisfy
\begin{align*}
    f_w(\omega_S, \omega_{T_i}) \geq \tau \geq f_w(\omega_S, \omega_{T_j}), \quad  \forall  \left( \omega_{T_i}, \omega_{T_j} \right) \in C_{\omega_S} \times \Omega(O_T)\backslash C_{\omega_S}.
\end{align*}

For example, as illustrated in Fig. \ref{Fig:Candidateexample}, when considering the term $\omega_{NCIT}=$ \textit{clear cell sarcoma of soft tissue} with $\tau=0.75$, the resulting subset $C_{\omega_{NCIT}}$ is the following
\begin{align*}
C_{\omega_{NCIT}} =\{adult \ soft \ part \ clear \ cell \ sarcoma, \ clear \ cell \ sarcoma, \ clear \ cell \ chondrosarcoma\}.
\end{align*}

Given a source entity $e_{S}$ described by $n$ labels, $n =|\Omega(e_{S})|$, MILA generates n subsets $C_{\omega_{S}}^{i}$, one for each label $\omega_S^{i} \in \Omega(e_S)$, with $i=1,2,...,n$. Let $C_{e_S}$ be the union of all generated subsets $C_{\omega_{S}}^{i}$:
\begin{align*} 
C_{e_S}=\bigcup_{i \in \{1,2,..,n\}} C_{\omega_{S}}^{i}.
\end{align*}

The function $f_{e}$ computes the similarity score between a source entity $e_{S} \in O_{S}$ and each target entity $e_{T} \in O_{T}$ verifying $C_{e_S} \cap \Omega(e_T) \neq \varnothing$ by calculating the maximum value of the similarity scores between each source label $\omega_{S} \in \Omega(e_{S})$ and each target label $\omega_{T} \in C_{e_S} \cap \Omega(e_T) $:
\begin{align*}
f_{e}(e_{S}, e_{T}) = \max \left(f_{w}(\omega_{S},\omega_{T}) \mid \omega_{S} \in \Omega(e_{S}), \omega_{T} \in C_{e_S} \cap \Omega(e_T) \right).
\end{align*}

Therefore, using the function $f_{e}$, MILA generates the set of the $m:= |C_{e_S}|$  most promising target correspondence entities for each source entity $e_{S}$: 
\begin{align*} C_{e_{S}} = \{e_{T_1}, e_{T_2}, ..., e_{T_m}\} \subseteq O_{T}, 0 \leq m \leq n*k
\end{align*}
verifying
\begin{align*}
    f_e(e_S, e_{T_i}) \geq \tau \geq f_e(e_S, e_{T_j}), \quad
    \forall  \left( e_{T_i},e_{T_j} \right) \in C_{e_S} \times O_T\backslash C_{e_S}.
\end{align*}

The set of \textit{source correspondence candidates} $C_{e_{S}}$ is transformed into an ordered sequence of elements $L_{e_S}=[e_{T_{i_1}}, e_{T_{i_2}} , ..., e_{T_{i_m}}] \subseteq O_{T}$ such that
\begin{align*}
    f_e(e_S, e_{T_{i_1}}) \geq f_e(e_S, e_{T_{i_2}}) \geq ...  \geq f_e(e_S, e_{T_{i_m}})
\end{align*}
where $\{{i_1},{i_2}, ... {i_m}\}$ is the permutation of indices that sorts the $C_{e_S}$ elements by the values of the function $f_e(e_S, e_T)$. All sequences $L_{e_S}$ are stored in the \textit{Source2Target Database} in Fig. \ref{Fig:overview}.
Similarly, for each target entity $e_{T}$, MILA generates the set of the most promising source entities  $C_{e_{T}}$, and the corresponding ordered sequence of elements, $L_{e_{T}}$.

For example, in the table on the right side of Fig. \ref{Fig:Candidateexample}, the most promising correspondence candidates for \textit{ncit:C3745} are listed along with their similarity scores, $f_e$. Although the entity \textit{DOID\_4880} is described by four labels in the ontology, only two of them have similarity scores that exceed $\tau$, \textit{childhood kidney clear cell sarcoma} and \textit{kidney clear cell sarcoma}. The maximum similarity value between these labels and some label of ncit:C3745 is achieved between the NCIT label \textit{clear cell sarcoma - not kidney} and the DOID label \textit{childhood kidney clear cell sarcoma}. Specifically, the value is 0.95621. Therefore, the similarity between \textit{ncit:C3745} and \textit{DOID\_4880} is set to 0.95621.

\subsection{The \textit{Retrieve-Identify-Prompt} pipeline}
In this subsection, we first define the two types of correspondences identified by MILA. We then outline the design of the prompt template used to query the LLM. Finally, we provide a detailed description of the algorithm.

\subsubsection{Types of correspondences}\label{identify}
Given a source entity $e_S \in O_S$ and a target entity $e_T \in O_T$, a \textit{bidirectional correspondence} between them exists if the following condition holds:
\begin{align*}e_{S} \in C_{e_{T}} \wedge e_{T} \in C_{e_{S}}.
\end{align*}
In other words, both entities must appear in each other’s candidate sets, allowing traversal in both directions. This property aligns with the fundamental principle of \textit{symmetry} in equivalence relations. Thus, for the equivalence relation $e_{S} \equiv e_{T}$ to hold, it is necessary that $e_{S}$ and $e_{T}$ are mutually accessible within their respective candidate sets.

It is important to note that the vector KBs index only the top-k most promising candidates. As a result, while a correspondence from an entity $e_S$ to another entity $e_T$ may be retrieved, the reverse correspondence from $e_T$ to $e_S$ is not always guaranteed. This limitation arises as the KB may prioritize other, more relevant candidates over $e_T$. Therefore, even if $e_{T} \in C_{e_{S}}$ it does not necessarily imply that $e_{S} \in C_{e_{T}} $ and vice versa.

Given a \textit{bidirectional correspondence} between a source entity $e_S \in O_S $ and a target entity $e_T \in O_T$, a \textit{high confidence bidirectional (HCB) correspondence} between both entities exists if they are the most prioritized entities in each other’s correspondence candidate sets:
\begin{align*} 
f_e(e_{S}, e_T)=\max \left (f_e(e_{S_i}, e_T) \mid e_{S_i} \in C_{e_T} \right)=\max \left(f_e(e_{S}, e_{T_j}) \mid  e_{T_j} \in C_{e_S} \right).
\end{align*}

\subsubsection{Prompt Template Design}\label{prompt}
In order to query an LLM, MILA uses a simple structured prompt with minimal ontology context to ensure clarity and focus \citep{peeters2023using}. The prompt includes only the ontology names and the preferred names of the source and target entities involved in the correspondence. This choice is based on studies reporting that a zero-shot scenario is almost as effective as a few-shot setting \citep{hertling2023olala, wang2023exploring}. The prompt is specifically structured to facilitate the LLM's binary decision-making. The prompt template is as follows:

\begin{center}
\fbox{\begin{minipage}{40em}
  You are a helpful expert in ontology matching, which involves determining equivalence correspondences between concepts from different ontologies. The source ontology is called $[src\_onto\_name]$ and the target ontology is called $[tgt\_onto\_name]$.

Classify whether the following concepts are equivalent:

Source concept: $[source\_entity]$

Target concept: $[target\_entity]$

If so, answer 'Yes', without adding any type of explanation. Otherwise, answer 'No'.
\end{minipage}}
\end{center}

\subsubsection{The algorithm}
The \textit{retrieve-identify-prompt} pipeline, as outlined in Algorithm \ref{alg:rip}, is designed to identify valid mappings between two ontologies. For each source entity $e_S$, the algorithm first retrieves the ordered sequence of target candidates $L_{e_S}$. Then iterates over this sequence using a PDFS strategy, looking for the first target entity $e_T \in L_{e_S}$ that is a valid match to $e_S$. Specifically, for each pair ($e_S,e_T)$, the algorithm checks whether the pair is a bidirectional correspondence. If so, it further verifies whether the pair is an HCB correspondence; if it is, the pair ($e_S,e_T)$ is added to the final mapping. If the pair is not an HCB correspondence, the algorithm queries an LLM to confirm the potential mapping. If the LLM confirms the mapping, it is added to the final mapping, and the search stops for that source entity. If the LLM does not confirm the potential mapping, the search continues with the next candidate in $L_{e_S}$, iterating until a valid mapping is identified or all candidates are exhausted.

\begin{algorithm}[ht]
\caption{Retrieve-Identify-Prompt pipeline}\label{alg:rip}
\begin{algorithmic}[1]
\State \textbf{Input:} $S$ (source\_entities), $LLM$, $PT$ (prompting\_template)
\State \textbf{Output:} $M$ (mapping)
\State Initialize $M$  $\gets$ \{\}
\ForAll {$ e_S \in S$}
    \State $L_{e_S}$ $\gets \textbf{retrieve}(e_S)$
    \State not\_found $\gets$ True
    \While{not\_found and $L_{e_S}$ is not empty} \Comment{\textbf{PDFS strategy}}
        \State $e_T$ $\gets$ pop($L_{e_S}$)
        \State $L_{e_T}$ $\gets \textbf{retrieve}(e_T)$ \Comment{\textbf{1. Retrieve}}
        \If{($e_S$,$e_T$) is a bidirectional correspondence} \Comment{\textbf{2. Identify}}
            \If{($e_S$,$e_T$) is an HCB correspondence}
                \State $M \gets M \cup \{(e_S,e_T)\}$
                \State not\_found $\gets$ False
            \Else
                \State $LLM\_answer \gets \textbf{prompt}(e_S,e_T, LLM, PT)$ \Comment{\textbf{3. Prompt}}
                \If {$LLM\_answer$ is 'Yes'}
                    \State $M \gets M \cup \{(e_S,e_T)\}$
                    \State not\_found $\gets$ False   
                \EndIf    
            \EndIf
        \EndIf
    \EndWhile
\EndFor
\end{algorithmic}
\end{algorithm}

\subsection{Examples}\label{examples}
This subsection presents two representative examples illustrating the \textit{retrieve-identify-prompt} pipeline. In the first example, MILA identifies an HCB correspondence in the first iteration of the pipeline, while in the second example, it iteratively applies the \textit{retrieve-identify-prompt} pipeline until a valid correspondence is found. In both cases, the source ontology is the NCI Thesaurus (NCIT) \citep{coronado2004nci} and the target ontology is the Disease Ontology (DOID) \citep{schriml2012disease}.

\subsubsection{Example 1}\label{example-1}
\begin{figure}[ht]
\centering
\includegraphics[width=0.8\linewidth]{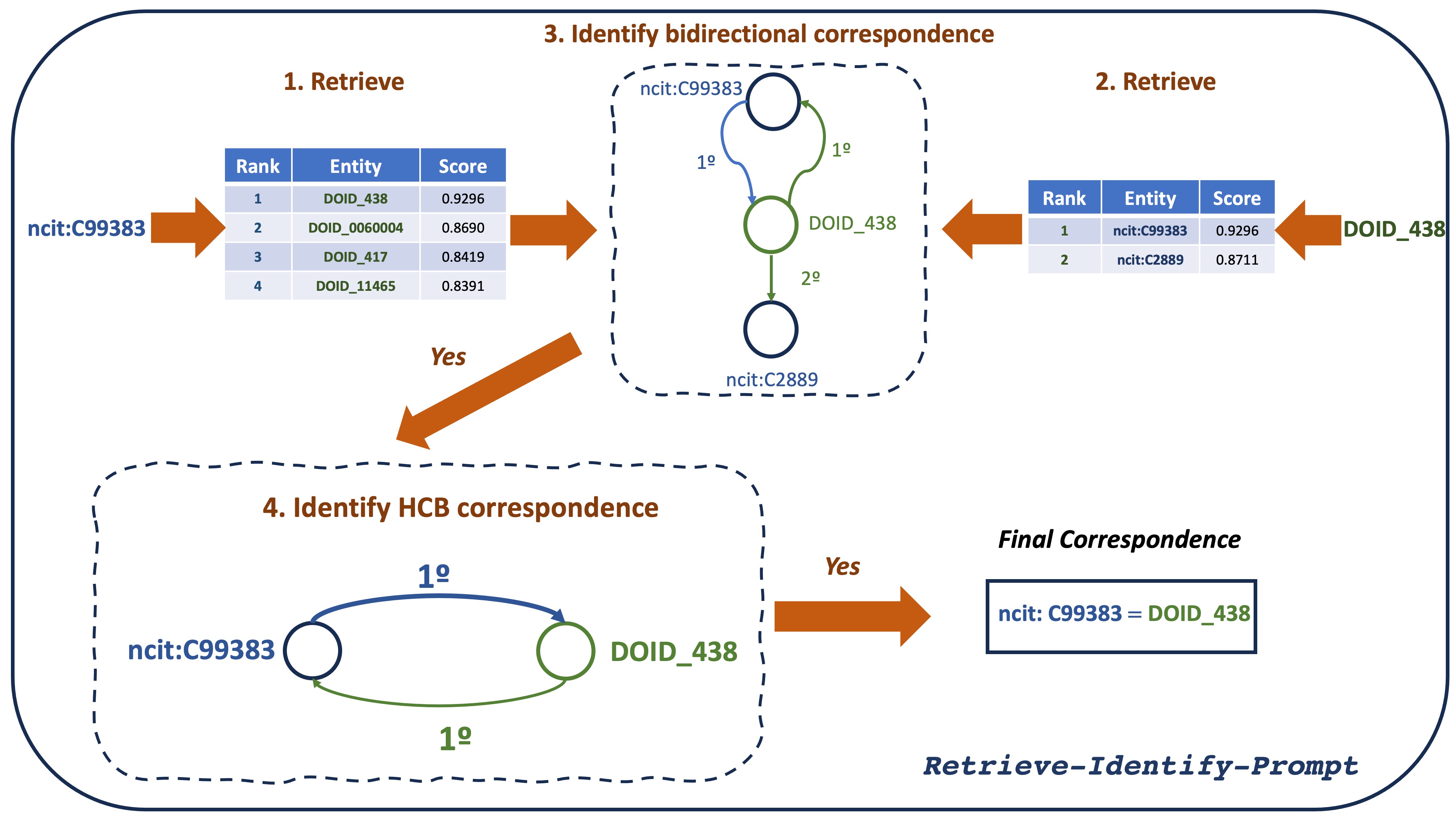}
\caption{Example showing the identification of an HCB correspondence between the source entity ncit:C99383 and the target entity DOID$\_438$.}\label{example1}
\end{figure}

Fig. \ref{example1} illustrates the alignment of the source entity ncit:C99383 to a corresponding entity in DOID, without requiring an LLM query, as an HCB correspondence is identified. In this case, ncit:C99383 is uniquely labeled as \textit{autoimmune nervous system disorder}.
\begin{itemize}
    \item \textbf{Step 1 - Retrieve}: MILA first retrieves the ordered sequence of target candidates, $L_{ncit:C99383}$, for the entity ncit:C99383. These candidates are the most promising target candidates for the label \textit{autoimmune nervous system disorder}: 
    \begin{enumerate}
    \item DOID:438 (autoimmune disease of the nervous system)
    \item DOID:0060004 (autoimmune disease of central nervous system)
    \item DOID:417 (autoimmune disease)
    \item DOID:11465 (autonomic nervous system disease)
    \end{enumerate}
    \item \textbf{Step 2 - Retrieve}: MILA retrieves the ordered sequence of the source candidates for the highest-ranked candidate in $L_{ncit:C99383}$, DOID:438.
    \item \textbf{Step 3 - Identify bidirectional correspondence}: Next, MILA checks whether there is a bidirectional correspondence between the highest ranked candidate, DOID:438, and the source entity. 
    \item \textbf{Step 4 - Identify HCB correspondence}: As there is a bidirectional correspondence, MILA checks if it is an HCB correspondence. Since ncit:C99383 and DOID:438 are each other's top-ranking candidates, MILA confirms the HCB correspondence and adds it to the set of final mappings: 

\begin{center}
\textit{O\textsubscript{NCIT} : autoimmune nervous system disorder  \( \equiv \) O\textsubscript{DOID} : autoimmune disease of the nervous system}
\end{center}
\end{itemize}

\subsubsection{Example 2}\label{example-2}

\begin{figure}[ht]
\centering
\includegraphics[width=\linewidth]{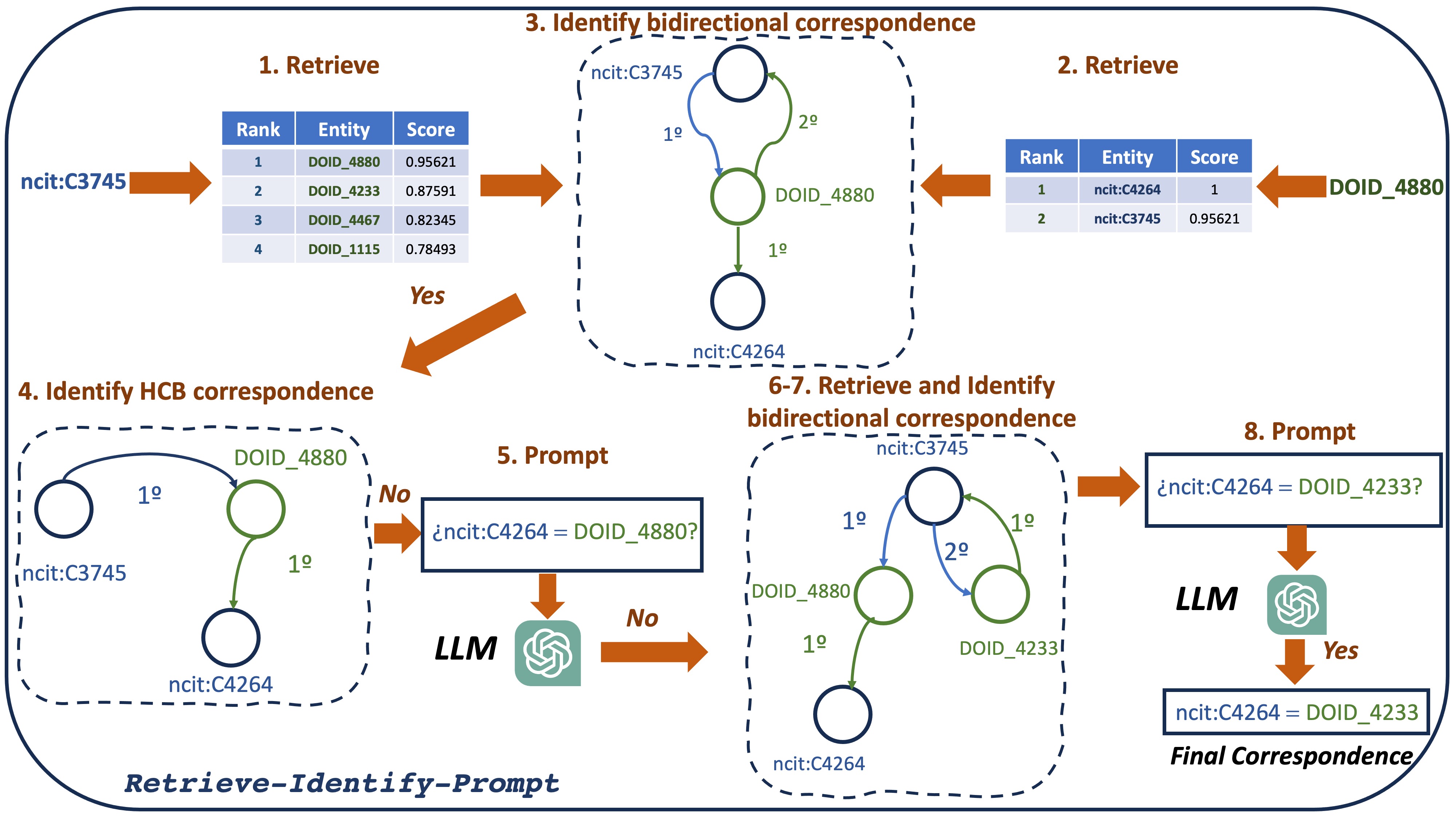}
\caption{Example showing the \textit{retrieve-identify-prompt} pipeline embedded into the PDFS strategy.} \label{example2}
\end{figure}

Fig. \ref{example2} illustrates how the source entity ncit:C3745 (\textit{clear cell sarcoma of soft tissue}) aligns with the target entity DOID:4233, requiring only two queries to the LLM - compared to the five queries needed by the baseline pipeline.
\begin{itemize}
    \item \textbf{Step 1 - Retrieve}: MILA first retrieves the ordered sequence of target candidates $L_{ncit:C3745}$. The left side of Fig. \ref{example2} shows only the four most promising of these candidates:
    \begin{enumerate}
    \item DOID:4880 (kidney clear cell sarcoma)
    \item DOID:4233 (clear cell sarcoma)
    \item DOID:4467 (renal clear cell carcinoma)
    \item DOID:1115 (sarcoma)
    \end{enumerate}
    \item \textbf{Step 2 - Retrieve}: MILA retrieves the ordered sequence of source candidates for the most promising target entity, $L_{DOID:4880}$.
    \item \textbf{Step 3 - Identify bidirectional correspondence}: Next, MILA checks whether there is a bidirectional correspondence between ncit:C3745 and DOID:4880.
    \item \textbf{Step 4 - Identify HCB correspondence}: As there is a bidirectional correspondence, it checks whether this is an HCB correspondence, by verifying whether ncit:C3745 is the top-ranked candidate for DOID:4880.
    \item \textbf{Step 5 - Prompt}: Since there is no bidirectional correspondence between ncit:C3745 and DOID:4880, MILA prompts the LLM to confirm the match. Since the LLM does not confirm the match between ncit:C3745 and DOID:4880, MILA proceeds to the next most similar candidate, DOID:4233 (clear cell sarcoma), and repeats the process.
    \item \textbf{Steps 6 and 7 - Retrieve and Identify bidirectional correspondence}: MILA retrieves the ordered sequence of source candidates for the second most promising target entity, DOID:4233, and then verifies that the pair (ncit:C3745, DOID:4233) is a bidirectional correspondence, but not an HCB correspondence.
    \item \textbf{Step 8 - Prompt}: MILA prompts the LLM to confirm the pair (ncit:C3745, DOID:4233). This time, the LLM confirms the match, allowing MILA to finalize the alignment and return the valid correspondence.
\end{itemize}

\section{OM Evaluation}\label{eval}
In this section, we report the experimental work we have carried out with the biomedical evaluation benchmark proposed by the OEAI in the 2024 edition \citep{jimenez2025proceedings}. In addition to its relevance, the biomedical domain is characterized by its terminology richness, which can be automatically processed by pre-trained language models and LLMs. Currently, the OAEI offers large OM datasets that cover the high variability in term expression typical of this domain \citep{kolyvakis2018biomedical}. In particular, the low performance achieved by LLM-based solutions in this domain encourages the development of new approaches \citep{giglou2024llms4om}. To further assess the robustness and broader applicability of MILA, we also report experimental results on other well-known tracks, including the Anatomy (mouse-human task) and Biodiversity (envo-sweet task) benchmarks.
\subsection{Evaluation Metrics}
All OM systems were evaluated using the traditional information retrieval metrics of Precision (P), Recall (R) and F-Measure. Precision reflects the correctness of an alignment achieved by an OM system ($\mathcal{A}$) with respect to a reference alignment ($\mathcal{R}$). It is usually defined as the ratio between the number of matches correctly detected by the OM system and the total number of matches identified by the OM system. Recall reflects the completeness of an alignment achieved by an OM system ($\mathcal{A}$) with respect to a reference alignment ($\mathcal{R}$). It is usually defined as the ratio of the number of matches correctly detected by the OM system to the total number of matches identified in the reference alignment. F-Measure combines P and R in a unique measurement.

\begin{align*} 
P\mathcal{(A,R)} &= \mathcal{\frac{|A \cap R|}{|A|}}  & R\mathcal{(A,R)}&= \mathcal{\frac{|A \cap R|}{|R|}}  & F-Measure\mathcal{(A,R)} &= 2*\frac{P*R}{P+R}
\end{align*}

\subsection{Dataset and Tasks}

The datasets used in our experiments cover an anatomy task, a biodiversity task, and five biomedical tasks. This diverse selection of datasets ensures a comprehensive evaluation of the proposed approach across multiple domains, highlighting its adaptability to different challenges of OM. 
\begin{itemize}
\item Anatomy OM Task: This real-world scenario involves aligning the Adult Mouse Anatomy ontology with the part of the NCI Thesaurus (NCIT) that describes human anatomy.
\item Biodiversity OM Task: This task focuses on finding correspondences between the Environment Ontology (ENVO) and the Semantic Web for Earth and Environment Technology (SWEET) Ontology, facilitating interoperability in environmental sciences.
\item Biomedical OM Tasks: These tasks involve ontology matching across six widely used biomedical ontologies: Online Mendelian Inheritance in Man (OMIM) \citep{hamosh2000online}, Orphanet Rare Disease Ontology (ORDO) \citep{vasant2014ordo}, SNOMED CT \citep{benson2021snomed}, Foundational Model of Anatomy (FMA) \citep{rosse2003reference}, Disease Ontology (DOID) \citep{schriml2012disease} and NCI Thesaurus (NCIT) \citep{coronado2004nci}.
\end{itemize}

Each task in the biomedical benchmark includes both equivalence matching and subsumption matching, although our experiments are focused only on equivalence OM. The quality of reference mappings is ensured both by human curation and by the use of several automated techniques, such as ontology pruning or enrichment with locality modules \citep{jimenez2011logmap,jimenez2017logmap}. The benchmark provides two different sets of test mappings \citep{he2022machine}. There is one set for unsupervised OM systems that do not use training mappings and another set for so-called semi-supervised systems, i.e. those that do use training mappings. Specifically, the first set contains the full set of reference mappings, while the second test set includes 70$\%$ of the set of reference mappings (excluding 30$\%$ of the training mappings). Even if an OM system is an unsupervised approach, such as MILA, it can report performance on this second set for comparison with the supervised OM systems.
   
\subsection{Final Configuration}
For all OM tasks, MILA used the SBERT retriever model \citep{reimers2019sentence}, which was set to multi-qa-distilbert-cos-v1. This choice was guided by the findings of the ablation study reported in \citep{hertling2023olala}, where this model achieved the highest recall at k=1, highlighting its outstanding ability to identify the most relevant correspondences — a crucial factor for accurately identifying HCB correspondences. Although the model did not achieve the best recall at k=5, it still provides a solid foundation for retrieving high-quality matches, making it a reliable choice for this task. Following \citep{giglou2024llms4om}, the value k during the search for the top k neighbors ($top_{k}$) was set to 5, as a compromise between the number of candidates to be generated and the recovery to be achieved. The candidates generated by SBERT with a confidence score lower than 0.75 were discarded and the scores were rounded to five decimal places. On the other hand, owlready2 \citep{lamy2017owlready} is used to extract information from ontologies, such as parent and child entities, textual labels, and annotations.

For language model-based processing, MILA used LLaMa-3.1 (70B) \citep{dubey2024llama}. The experiments were conducted on a desktop computer equipped with an Intel Core i5 (4 cores, 3.50 GHz) processor, 32 GB RAM and an AMD Radeon R9 M295X (4GB) graphic card. LLaMa-3.1 (70B) was executed via an inference endpoint, leveraging 4× NVIDIA L40S GPUs. All LLM parameters were kept at their default values, with a temperature of 0.7 to maintain a balance between creativity and logical coherence. Each experiment using LLaMa-3.1 (70B) was executed in a minimum of 20 executions to obtain a reliable estimate of variance. Although a higher number of executions per experiment would provide a more stable and accurate measure of variability \citep{giglou2024llms4om}, the high associated costs prevented this approach. For runtime comparisons between the baseline and MILA pipelines, MILA used LLaMa-3.3 (8B), as it could be executed on our desktop computer. Running LLaMa-3.1 (70B) via an inference endpoint for the baseline pipeline was unaffordable due to the associated high costs.
\subsection{Experimental Results on the biomedical evaluation benchmark}
In this study, we compared the performance of MILA to fourteen OM systems on the evaluation benchmark: AMD \citep{wang2022amd}, BERTMap \citep{he2022bertmap}, BERTMapLt \citep{he2023deeponto}, BioGITOM \citep{BioGITOM}, BioSTransMatch \citep{BioSTransMatch}, HybridOM \citep{HybridOM}, LLMs4OM \citep{giglou2024llms4om}, LogMap \citep{jimenez2011logmap}, LogMapBio \citep{jimenez2017logmap}, LogMapLt  \citep{jimenez2017logmap}, Matcha \citep{faria2023results}, Matcha-DL \citep{faria2023results}, OLaLa \citep{hertling2023olala} and SORBETMatcher  \citep{gosselin2023sorbet}. Data were compiled from results published in the OAEI BIO-ML track (2023 and 2024 editions), as well as relevant literature, with no OM system intentionally excluded to the best of our knowledge. This comparison highlights the strengths and limitations of current systems, providing a benchmark to evaluate the improvements achieved by MILA in terms of performance and computational overhead. 

In short, some state-of-the-art OM systems (e.g., BertMap, Matcha, Olala or MILA) use only textual knowledge to predict mapping candidates, whereas others (e.g., LogMap, SORBETMatcher or LLMs4OM) also use structure knowledge. Additionally, most OM systems use pre-trained neuronal models (such as SBERT or BERT) to encode ontology entities. In some cases, they also include a fine-tuning stage (e.g., BertMap or Matcha-DL), or domain-specific knowledge (e.g., BioGITOM, BioSTransMatch or LogMapBio). Finally, mapping refinement is mainly based on heuristic and logical reasoning, with the exception of Olala, LLMs4OM and MILA, which leverage LLMs for mapping refinement. Compared to other current approaches, MILA's main distinction is its use of a PDFS strategy, supported by LLMs, to solve mapping refinement. For more detailed information on the characteristics of comparative systems, the reader is referred to \ref{app3}.
\subsubsection{Results in the Unsupervised Setting}

Tables \ref{tab:MondoUnsupResults} and \ref{tab:UMLSUnsupResults} show the performance of all OM systems in the unsupervised setting of the evaluation dataset. MILA is the best performing algorithm in four of the five OM tasks. Even for the task in which it does not achieve the best results, its performance is comparable to that of the leading system. Below we provide a detailed interpretation of these results for each of the evaluated tasks.

\begin{itemize}
    \item In the \textbf{OMIM-ORDO mapping task}, MILA's F-Measure score is outstanding compared to the other approaches. It outperforms the second best OM system in this task, LogMapBio, by 17\%. Specifically, MILA achieves high recall, compared to the rest of the approaches, due to the high recall of both the retrieval module and the \textit{retrieve-identify-prompt} pipeline. Compared to the other tasks, the OMIM-ORDO task achieves the lowest F-Measure. 
    \item In the \textbf{NCIT-DOID mapping task}, MILA outperforms the second best baseline, HybridOM, in terms of F-Measure by $3\%$. In addition, LogMapBio and SORBETMatcher also achieve similar results. In short, this is the task with the highest average F-Measure score of all approaches. 
    \item In the \textbf{SNOMED-FMA mapping task}, MILA's F-Measure score is outstanding compared to the other approaches. It outperforms the second-best OM systems in this task, BERTMap and HybridOM, by 13\%. Again, MILA achieves a high recall, compared to the rest of the approaches, and a precision comparable to the best proposals (family BERTMap). 
    \item In the \textbf{SNOMED-NCIT (Pharmacology) mapping task}, MILA is the second-best approach with $4\%$ below the outstanding approach, HybridOM, in terms of F-Measure. Note that MILA outperforms the third baseline (i.e. AMD) in terms of F-Measure by $9\%$. 
    \item In the \textbf{SNOMED-NCIT (Neoplasm) mapping task}, MILA's F-Measure score is also outstanding compared to the other approaches. It outperforms the second best OM system in this task, LogMapBio, by 17$\%$. As in the other tasks, MILA achieves high recall, compared to the rest of the approaches. 
\end{itemize}

\begin{table}
\begin{center}
  \begin{tabular}{|c|c|c|c|c|c|c|}
    \hline
    \multirow{2}{*}{\textbf{OM System}} &
      \multicolumn{3}{c|}{\textbf{OMIM-ORDO}} &
      \multicolumn{3}{c|}{\textbf{NCIT-DOID}} \\
    & P & R & F-Measure& P & R & F-Measure \\
    \hline
    AMD & 0.664&0.508&0.576 & 0.885&0.691&0.777 \\
    \hline
    BERTMap &0.734&0.576&0.646&0.888&0.878&0.883\\
    \hline
    BERTMapLt & 0.834&	0.497&	0.623&0.919&0.772&0.839\\
    \hline
    BioSTransMatch & 0.312	&0.586	&0.407 & 0.657&0.833&0.735\\
    \hline
    HybridOM & 0.690&	0.679	&0.685 & 0.924&0.913&0.918\\
    \hline
    LLMs4OM & 0.718&	0.580	&0.641& 0.862&0.801 & 0.830\\ 
    \hline
    LogMap & 0.876&	0.448&	0.593 & 0.934&0.668&0.779\\
    \hline
    LogMapBio & 0.866&	0.609&	0.715& 0.860&\textbf{0.962}&0.908 \\
    \hline
    LogMapLt & \textbf{0.940}&	0.252&	0.397 &\textbf{0.983}&	0.575&	0.725\\
    \hline
    Matcha & 0.781	&0.509	&0.617& 0.882&	0.756&	0.814\\
    \hline
    Matcha-DL & 0.745&	0.513&	0.607 & 0.847&	0.586&	0.693\\
    \hline
    OLaLa & 0.735&	0.582&	0.649 & 0.913&	0.864	&0.888\\
    \hline
    SORBETMatcher &0.693&	0.635	&0.663 & 0.920	&0.907&	0.913\\
    \hline
    MILA & 0.879	&\textbf{0.778}	&\textbf{0.831}& 0.964&	0.932&	\textbf{0.948}\\ \hline

  \end{tabular}
  \caption{Performance comparison for the the OMIM-ORDO and NCIT-DOID tasks in the unsupervised setting.}
  \label{tab:MondoUnsupResults}
\end{center}  
\end{table}

\begin{table}
\begin{center}
  \begin{tabular}{|c|c|c|c|c|c|c|c|c|c|}
    \hline
    \multirow{2}{*}{\textbf{OM System} }&
      \multicolumn{3}{c|}{\textbf{SNOMED-FMA}} &
      \multicolumn{3}{c|}{\textbf{SNOMED-NCIT (Pharm)}} &
      \multicolumn{3}{c|}{\textbf{SNOMED-NCIT (Neopl)}} \\
    & P & R & F-Measure& P& R& F-Measure & P & R & F-Measure\\
    \hline
    AMD &0.890&0.633&0.740&0.962&0.670&0.790&0.836&0.481&0.610\\
    \hline
    BERTMap &\textbf{0.979}&0.662&0.790& 0.971&	0.585&	0.730& 0.557	&0.762	&0.643\\
    \hline
    BERTMapLt&\textbf{0.979}&	0.655&	0.785& 0.981&	0.574	&0.724&0.831&	0.687	&0.752\\
    \hline
    BioSTransMatch &0.128&	0.384&	0.192&0.584&0.844&	0.690&0.289	&0.663&	0.402\\
    \hline
    HybridOM &0.870&0.722&	0.790&0.916&	\textbf{0.889}&	\textbf{0.902}&0.807&	0.710&	0.755\\
    \hline
    LLMs4OM &0.211&0.326&0.256&0.818&0.582&0.680&0.470	&0.530	&0.495\\ 
    \hline
    LogMap &0.744&	0.407&	0.526&0.966&	0.607&	0.746&0.870&	0.586&	0.701\\
    \hline
    LogMapBio&0.827&0.577&	0.680&0.928&	0.611&	0.737&0.748	&0.795	&0.771  \\
    \hline
    LogMapLt &0.970&0.542&	0.696&\textbf{0.996}	&0.599&0.748&\textbf{0.951}	&0.517&	0.670\\
    \hline
    Matcha&0.887&	0.502&	0.641&0.987	&0.607&	0.752&0.838	&0.551&	0.665 \\
    \hline
    Matcha-DL&0.960&	0.602&	0.740&0.904	&0.616&	0.733&0.811&	0.514&	0.629\\
    \hline
    OLaLa &0.270&	0.348&	0.304&---&---&---&0.540	&0.546&	0.543\\
    \hline
    SORBETMatcher&0.618&0.749&	0.677&0.973	&0.607	&0.748&0.626	&0.642	&0.634 \\
    \hline
    MILA &0.964&\textbf{0.834}	&\textbf{0.894}&0.981&0.772&	0.864&0.928&	\textbf{0.880}&\textbf{	0.903}\\ \hline
    
  \end{tabular}
  \caption{Performance comparison for the SNOMED-FMA (Body), SNOMED-NCIT (Pharmacology) and SNOMED-NCIT (Neoplasm) tasks in the unsupervised setting.}
  \label{tab:UMLSUnsupResults}
\end{center}
\end{table}

\subsubsection{Results in the Semi-Supervised Setting}

Although MILA is an unsupervised system, we also show its performance against systems that use data training to improve their results. In the semi-supervised setting, MILA is the best performing algorithm in two of the five ontology mapping tasks (see Tables \ref{tab:MondosupResults} and \ref{tab:UMLSsupResults}). Specifically, in the NCIT-DOID task, MILA outperforms the second-best OM system in this task, BioGITOM, by 6$\%$, achieving an F-Measure of 0.97. In the SNOMED-NCIT task, it outperforms the second-best OM system, Matcha-DL, by 18$\%$. In addition, for the rest of the tasks, MILA is the second or third best approach with an F-Measure between 3$\%$ and 4$\%$ below the leading approaches. Please note that no information is available for LLMs4OM, and therefore it does not appear in the tables.

\begin{table}
\begin{center}
  \begin{tabular}{|c|c|c|c|c|c|c|}
    \hline
    \multirow{2}{*}{\textbf{OM System}} &
      \multicolumn{3}{c|}{\textbf{OMIM-ORDO}} &
      \multicolumn{3}{c|}{\textbf{NCIT-DOID}} \\
    & P & R & F-Measure& P & R & F-Measure \\
    \hline
    AMD & 0.601&0.567&	0.583 & 0.858&0.770&0.811 \\
    \hline
    BERTMap &0.645&0.592&0.617&0.831&	0.883&0.856\\
    \hline
    BERTMapLt & 0.782&0.507&0.615&0.888&0.770&0.825\\
    \hline
    BioGITOM&0.951&0.773&\textbf{0.853}&0.944&0.884&0.913\\
    \hline
    BioSTransMatch & \textbf{0.973}&0.278&0.432 & 0.698&0.741&0.719\\
    \hline
    HybridOM & 0.611&0.683&0.645 & 0.895&0.913&0.904\\
    \hline
    LogMap & 0.834&0.456&0.589& 0.908&0.664&0.767\\
    \hline
    LogMapBio & 0.821&0.614&0.703& 0.811&\textbf{0.959}&0.879 \\
    \hline
    LogMapLt & 0.919&0.261&0.407 &\textbf{0.976}&0.575&0.723\\
    \hline
    Matcha & 0.718&0.519&0.602& 0.839&0.750&0.792\\
    \hline
    Matcha-DL & 0.745&0.732&0.738& 0.847&0.834&0.841\\
    \hline
    OLaLa & 0.655&0.570&0.610 & 0.880&0.861&0.870\\
    \hline
    SORBETMatcher &0.568&0.652&0.607& 0.925&0.882&0.903\\
    \hline
    MILA & 0.874&\textbf{0.784}&0.827	&0.967&0.928&\textbf{0.970}\\ \hline

  \end{tabular}
  \caption{Performance comparison for the OMIM-ORDO and NCIT-DOID tasks in the semi-supervised setting.}
  \label{tab:MondosupResults}
\end{center} 
\end{table}

\begin{table}[ht]
\begin{center}
  \begin{tabular}{|c|c|c|c|c|c|c|c|c|c|}
    \hline
    \multirow{2}{*}{\textbf{OM System} }&
      \multicolumn{3}{c |}{\textbf{SNOMED-FMA}} &
      \multicolumn{3}{c |}{\textbf{SNOMED-NCIT (Pharm)}} &
      \multicolumn{3}{c|}{\textbf{SNOMED-NCIT (Neopl)}} \\
    & P & R & F-Measure& P& R & F-Measure & P & R & F-Measure\\
    \hline
    AMD &0.861&0.709&0.778&0.952&0.746&0.836&0.792&0.528&0.633\\
    \hline
    BERTMap &0.970&0.669&0.792&0.898&0.715&0.796&0.562&0.771&0.650\\
    \hline
    BERTMapLt&0.970&0.662&0.787&0.973&0.569&0.718&0.775&0.688&0.729	\\
    \hline
    BioGITOM&0.962&\textbf{0.886}&\textbf{0.923}&0.983&0.713&0.827&---&---&---\\
    \hline
    BioSTransMatch &0.357&0.661&0.464& 0.845&0.860&0.852&0.700&0.607&0.650\\
    \hline
    HybridOM &0.825&0.725&0.772&0.884&	\textbf{0.886}&0.885&0.747&0.718&0.732\\
    \hline
    LogMap &0.673&0.411	&0.511&0.952&0.603&0.738&0.823&0.583&0.683\\
    \hline
    LogMapBio&0.770&0.577&0.660&0.899&0.606&0.724&0.675&0.793&0.729\\
    \hline
    LogMapLt &0.958&0.542&0.693	& \textbf{0.994}	&0.594	&0.743&\textbf{0.931}&	0.514&0.662\\
    \hline
    Matcha&0.846	&0.502&	0.630&0.982&0.601&0.746 &0.782&0.545&0.642\\
    \hline
    Matcha-DL&0.959&0.825&0.887&0.903&0.872&\textbf{0.888}&0.806&0.714&0.757\\
    \hline
    OLaLa &0.202	&0.339	&0.253&---&---&---&0.451	&0.545	&0.493\\
    \hline
    SORBETMatcher&0.794&0.704&0.746 &0.876&0.604&0.715&0.731&0.605&0.662 \\
    \hline
    MILA &\textbf{0.967}&0.815&0.884&0.979&0.764&0.858&0.926&\textbf{0.863}&\textbf{0.893}\\ \hline
    
  \end{tabular}
  \caption{Performance comparison for the SNOMED-FMA (Body), SNOMED-NCIT (Pharmacology) and SNOMED-NCIT (Neoplasm) tasks in the semi-supervised setting.}
  \label{tab:UMLSsupResults}
\end{center}
\end{table}

\subsection{Experimental Results on the anatomy and biodiversity evaluation benchmarks}
In this section, we report the performance of MILA on the Anatomy evaluation benchmark (MOUSE-HUMAN task) and the Biodiversity evaluation benchmark (ENVO-SWEET task). First, we compared MILA with the four top performing systems from the OAEI 2024 Anatomy track: Matcha \citep{faria2023results}, MDMapper \citep{liu2024mdmapper}, LogMap \citep{jimenez2011logmap}, and LogMapBio \citep{jimenez2017logmap}. The evaluation results, compiled from the official benchmark dataset, are presented in Table \ref{tab:RestResults}. MILA achieved the second-best performance, with $2\%$ below the outstanding approach, Matcha, in terms of the F-Measure. Next, we assessed MILA's performance on the ENVO-SWEET task of the OAEI 2024 biodiversity track, where all competing systems belonged to the LogMap family \citep{jimenez2017logmap}. The results, summarized in Table \ref{tab:RestResults}, indicate that MILA outperforms all other approaches in this benchmark. In particular, it outperforms LogMap, the second best system, by  $17\%$.
\begin{table} [ht]
\begin{center}
  \begin{tabular}{|c|c|c|c|c|c|c|}
    \hline
    \multirow{2}{*}{\textbf{OM System}} &
      \multicolumn{3}{c|}{\textbf{MOUSE-HUMAN}} &
      \multicolumn{3}{c|}{\textbf{ENVO-SWEET}} \\
    & P & R & F-Measure& P & R & F-Measure \\
    \hline
    Matcha & 0.951&\textbf{0.931}&\textbf{0.941} & ---&---&--- \\
    \hline
    MDMapper &0.926	&0.881&0.903& ---&---&--- \\
    \hline
    LogMap &0.917&0.848&0.881&0.776&0.659&0.713\\
    \hline
    LogMapBio &0.888&0.898&0.908& ---&---&--- \\
    \hline
    LogMapKG& ---&---&---& 0.775&0.658&0.711\\
    \hline
    LogMapLt& 0.962&0.728&0.828& 0.803&0.595&0.683\\
    \hline
    MILA &\textbf{0.974}&0.875&0.922&\textbf{0.951}&\textbf{0.748}&\textbf{0.837}\\
    \hline    
  \end{tabular}
  \caption{Performance comparison for the the MOUSE-HUMAN and ENVO-SWEET tasks.}
  \label{tab:RestResults}
\end{center}  
\end{table}

\subsection{Runtime analysis}\label{runtime}
This section provides the runtimes of MILA for the datasets used in our experiments. Since MILA consists of three major steps, we report the runtimes for each of them: KB building, mapping prediction, and refinement (the \textit{retrieve-identify-prompt} pipeline). Table \ref{tab:KBBuilding} shows the runtimes associated with the KB building process, while \ref{tab:runtimes70B} presents the runtimes for mapping prediction and refinement.

\subsubsection{KB building}
Table \ref{tab:KBBuilding} shows the processing time required to build the source and target KBs with SBERT. This runtime increases with the number of labels in the ontology. This observation is consistent with theoretical expectations, since the time complexity of the KB construction, $O(n \cdot L^2)$, depends on both the total number $n$ of labels and the quadratic complexity of processing each label in transformer-based models - where $L$ denotes the average token length per label. Scalability challenges arise primarily when $L$ is large due to the quadratic cost of self-attention in transformers \cite{vaswani2017attention}. Therefore, ontologies with longer entity names, such as the pair OMIM-ORDO, show longer times compared to other pairs with higher number of labels, such as NCIT-DOID. However, since MILA only indexes entity names, which are usually short, this complexity remains tractable. As a result, KB construction in MILA scales efficiently, even for large ontologies, maintaining tractability across a variety of use cases.
\begin{table}[ht]
\begin{center}
\begin{tabular}{| c | c | c | c | c |}
    \hline
     & \textbf{Number of}& \textbf{Number of} & \textbf{Total number} &\textbf{KB building}\\ 
      \textbf{OM TasK} &\textbf{source labels} &\textbf{target labels} & \textbf{of labels} &  \textbf{time}\\ \hline
    OMIM-ORDO &25,890 & 21,556& 47,446&00:04:57\\
    NCIT-DOID &48,619 & 14,936&63,555&00:04:56\\
    SNOMED-FMA &44,927 & 142,984&187,911& 00:23:23\\
    SNOMED-NCIT (Pharm) &33,456 & 64,980&98,436&00:11:56\\
    SNOMED-NCIT (Neoplasm) &38,946 & 60,597&99,543&00:17:06 \\
    MOUSE-HUMAN &2,737 & 5,096& 7,833&00:00:35\\
    ENVO-SWEET &9,430 & 4,365& 13,795&00:01:02 \\ \hline
\end{tabular}
\caption{Runtime KB construction in MILA.}
\label{tab:KBBuilding}
\end{center}
\end{table}
\subsubsection{Mapping prediction}
Table \ref{tab:runtimes70B} shows that, in some cases, the processing time spent to predict correspondences can be significantly higher than the time required to build the KBs (see Table \ref{tab:KBBuilding}). For example, SNOMED-FMA, the largest dataset with 7,256 source labels in the reference dataset and 142,984 target labels, required more than 2 hours, which is more than six times the runtime of the first step (KB building). In contrast, for other tasks, such as NCIT-DOID, both stages (KB building KB and mapping prediction) were completed in approximately the same amount of time. This result is consistent with theoretical expectations, since the computation of similarities using cosine similarity between a query vector and all vectors stored in the target KB follows a complexity of $O(n_T \cdot d)$, where $n_T$ is the number of target labels and $d$ is the embedding dimension. Additionally, retrieving the top-$k$ most similar vectors adds a logarithmic term, $O(n_T \cdot d + n_T \cdot \log k)$. However, since $k$ is typically much smaller than $n_T$, the complexity simplifies to $O(n_T\cdot d)$ per query, leading to a total query complexity of $O(n_S \cdot n_T \cdot d)$, where $n_S$ is the number of source labels to match. Consequently, as the size of both ontologies grows, the time required increases at a rate proportional to their product, resulting in a quadratic time complexity of the correspondence prediction step. 

\begin{table}[ht]
\begin{center}
\begin{tabular}{| c | c | c | c | c | c |}
    \hline
     & \textbf{Number of} &\textbf{Number of}& \textbf{Mapping}& \textbf{\textit{Retrieve-identify-prompt}}&\textbf{Average}\\ 
      \textbf{OM} &\textbf{target} & \textbf{entities in}& \textbf{prediction}& \textbf{pipeline}&\textbf{Variance}\\ 
     \textbf{TasK}&\textbf{labels}& \textbf{reference}& \textbf{time} &  \textbf{time} &\textbf{in LLM}\\ \hline
    OMIM-ORDO & 21,556&3,721&00:21:32&00:12:41&0.0138\\
    NCIT-DOID & 14,936&4,686&00:26:39&00:04:45&0.0086\\
    SNOMED-FMA & 142,984& 7,256&02:08:34&00:12:35&0.0085\\
    SNOMED-NCIT (Pharm) &64,980&5,803&00:36:51&00:26:28&0.0817\\
    SNOMED-NCIT (Neopl) &60,597&3,804 &00:40:59&00:11:32&0.0106\\
    MOUSE-HUMAN &5,096& 1,516 &00:01:47&00:01:38&0.0274\\
    ENVO-SWEET &4,365&805 &00:02:02&00:00:49&0.0055\\ \hline
\end{tabular}
\caption{Runtime of Mapping Prediction and Refinement Using MILA with LLaMa-3.1-70B.}
\label{tab:runtimes70B}
\end{center}
\end{table}
\subsubsection{The retrieve-identify-prompt pipeline}
The runtime analysis in Table \ref{tab:runtimes70B} shows that the \textit{retrieve-identify-prompt} pipeline remains efficient, with execution times consistently below 13 minutes in most tasks, except for SNOMED-NCIT (Pharm). Although the reference dataset for SNOMED-NCIT (Pharm) contains fewer entities (5,803) compared to the largest task, SNOMED-FMA (7,256), the pipeline for SNOMED-NCIT (Pharm) takes more than twice as long as for SNOMED-FMA. This discrepancy is mainly attributed to MILA identifying a lower proportion of high-confidence bidirectional (HCB) matches in SNOMED-NCIT (Pharm), resulting in a higher number of LLM queries, which in turn increases the execution time. Furthermore, the low average variability in LLM responses across all tasks suggests stable performance, regardless of the size and complexity of the ontology.

Table \ref{tab:runtimeComparing} presents a comparison of the runtimes between the \textit{retrieve-identify-prompt} and \textit{retrieve-then-prompt} pipelines across various OM tasks, using LLaMa-3.3-8B-Instruct. The results show that the \textit{retrieve-identify-prompt} pipeline (column 3) requires more than three times the execution time when querying LLaMa-3.3-8B-Instruct compared to querying LLaMa-3.1-70B (column 5, Table \ref{tab:runtimes70B}). This discrepancy is mainly due to the fact that the LLaMa-3.3-8B-Instruct version operates on a machine with fewer computational resources. Furthermore, the \textit{retrieve-identify-prompt} pipeline significantly reduces execution time compared to the baseline pipeline, demonstrating how MILA minimizes unnecessary LLM queries. Even for larger ontologies such as SNOMED-FMA, MILA completes in approximately 45 minutes, while the baseline takes almost 47 hours. In the case of SNOMED-NCIT (Pharm), the differences in execution time are smaller since, as we have already mentioned, MILA identifies a lower proportion of HCB matches. These results highlight MILA's efficiency in reducing LLM queries and accelerating OM, particularly for large-scale datasets.

\begin{table}[ht]
\begin{center}
\begin{tabular}{| c | c | c | c |}
    \hline
     \textbf{OM} & \textbf{Number of entities} & \textbf{\textbf{\textit{Retrieve-identify-prompt}}}&\textbf{\textit{Retrieve-then-prompt}}\\ 
      \textbf{TasK}& \textbf{in reference dataset}& \textbf{Time}&\textbf{Time}\\ \hline
    OMIM-ORDO &3,721&00:43:51&44:53:05\\
    NCIT-DOID &4,686&00:20:07 &32:49:56\\
    SNOMED-FMA &7,256& 00:44:59 & 46:46:53\\
    SNOMED-NCIT (Pharm) &5,803&02:03:31&35:32:36\\
    SNOMED-NCIT (Neoplasm) &3,804&00:34:05 &26:11:24\\
    MOUSE-HUMAN &1,516& 00:11:49&09:27:16\\
    ENVO-SWEET &805& 00:04:10&04:13:19\\ \hline
\end{tabular}
\caption{Comparison of runtimes between the \textit{retrieve-identify-prompt} and \textit{retrieve-then-prompt} pipelines, using LLaMa-3.3-8B-Instruct.}
\label{tab:runtimeComparing}
\end{center}
\end{table}
These experimental results align with theoretical expectations. The time complexity in both baseline and MILA pipelines is $O(n \cdot k)$, where $n$ represents the number of source labels and $k$ the number of target candidates retrieved per source label. Therefore, the time complexity scales linearly with the number of source labels \citep{giglou2024llms4om}. However, MILA adds a layer of decision-making that saves time by reducing unnecessary LLM queries. First, MILA incorporates a step where the first candidate can be quickly identified as an HCB correspondence. This step involves simple heuristics that can be done in constant time $O(1)$, eliminating $k$ LLM queries for each entity for which an HCB correspondence is identified. Second, if the first candidate is not identified as an HCB correspondence, then MILA proceeds with the PDFS strategy, which can also minimize the number of queries by ending early when a valid match is found. In Example 2, MILA completes the alignment in just two queries to the LLM, whereas the basic pipeline requires three additional queries.

\subsubsection{Performance on using different configurations}
Finally, Table \ref{tab:MILAConfigResults} presents the performance scores for all tasks across different MILA configurations: MILA-HCB, which works without prompting any LLM and is based only on the HCB correspondences; MILA-LLaMa-3 (8B), which queries LLaMa-3.3 (8B-Instruct); and MILA-LLaMa-3 (70B), which leverages LLaMa-3.1 (70B). The results indicate that MILA-HCB consistently achieves high precision across all datasets but tends to have a lower recall. MILA-LLaMa-3 (70B) tends to provide superior F-Measure results. In some cases, MILA-LLaMa-3 (8B) equals or improves the F-Measure, although its impact varies depending on the dataset.  Although the gain achieved by MILA-LLaMa3-70B is not substantially higher compared to MILA-HCB (except for SNOMED-NCIT-Pharm), it is important to note that MILA-LLaMa3-70B is built on top of MILA-HCB to address the borderline cases where MILA-HCB fails to find a HCB correspondence. Therefore, the benefit of MILA-LLaMa3-70B is in enhancing the overall system's performance by solving the challenging mappings that MILA-HCB cannot resolve. 

\begin{table}[ht]
\begin{center}
  \begin{tabular}{|c|c|c|c|c|c|c|c|c|c|}
    \hline
    \multirow{2}{*}{\textbf{OM Test} }&
      \multicolumn{3}{c|}{\textbf{MILA-HCB}} &
      \multicolumn{3}{c|}{\textbf{MILA-LLaMa3-8B}} &
      \multicolumn{3}{c|}{\textbf{MILA-LLaMa3-70B}} \\
    & P & R & F-Measure& P& R& F-Measure & P & R & F-Measure\\
    \hline
    OMIM-ORDO &\textbf{0.911}&0.738&0.816&0.827&\textbf{0.782}&0.804&0.879&0.778&\textbf{0.831}\\
    \hline
    NCIT-DOID &\textbf{0.968}&0.907&0.936&0.943&0.929&0.936&0.964&\textbf{0.932}&\textbf{0.948}\\
    \hline
    SNOMED-FMA&\textbf{0.975}&0.799&0.878&0.951&\textbf{0.844}&0.894&0.964&0.834&\textbf{0.894}\\
    \hline
    SNOMED-NCIT(P)&\textbf{0.988}&0.625&0.766&0.958&\textbf{0.821}&\textbf{0.884}&0.891&0.772&0.864\\
    \hline
    SNOMED-NCIT(N) &\textbf{0.932}&0.802&0.862&0.907&0.873&0.890&0.928&\textbf{0.880}&\textbf{0.903}\\
    \hline
    MOUSE-HUMAN &\textbf{0.985}&0.835&0.904&0.920&0.859&0.889&0.974&\textbf{0.875}&\textbf{0.922}\\
    \hline
    ENVO-SWEET &\textbf{0.957}&0.718&0.820&0.935&\textbf{0.749}&0.832&0.951&0.748&\textbf{0.837}\\
    \hline
  \end{tabular}
  \caption{Performance using different MILA configurations.}
  \label{tab:MILAConfigResults}
\end{center}
\end{table}
\section{Discussion and Future Work}\label{conclusions}
OM plays a key role in enabling smooth communication between heterogeneous applications and ensuring data interoperability and integration across diverse domains. Despite its importance, OM remains an evolving field that continues to benefit from advances in machine learning and language modeling to improve matching performance and scalability. Our experimental results show that MILA significantly improves current LLM-based OM systems in both F-Measure performance and execution efficiency. By introducing an intermediate step that identifies HCB correspondences, MILA enhances significantly the F-Measure with regard the current LLM-based OM systems. Additionally, identification of such matches is performed in constant time, eliminating the need to perform $k$ LLM queries for each HCB correspondence. The PDFS strategy also minimizes the number of LLM queries by early ending when a valid match is found. Therefore, although time complexity remains consistent with current LLM-based OM systems, MILA offers substantial practical benefits by reducing the number of LLM queries, which can lead to faster matching for large-scale ontologies, as shown in the results achieved in our study. The exact efficiency gain depends on the quality of the retrieval system (how well it ranks relevant candidates), the distribution of correspondence candidates (how early in the retrieval list the correct correspondence typically appears), and the scalability of the LLM (how expensive each query is in terms of time). To achieve efficiency gains, MILA incorporates a retrieval system that prioritizes correspondences between entities with the most semantically similar synonyms. Therefore, although MILA maintains the same theoretical time complexity as current LLM-based OM systems, its ability to minimize unnecessary LLM interactions offers substantial practical benefits, as evidenced by the results in our study.

Moreover, MILA shows a substantial improvement over state-of-the-art OM systems, in terms of its F-Measure performance. First, our work reports that MILA outperforms leading OM systems in four of the five tasks in the unsupervised setting, and in two of the five tasks in the semi-supervised setting, despite being a zero-shot approach. Second, our approach exhibits task-agnostic performance, remaining nearly stable across all tasks and settings, unlike other approaches, whose performance tends to vary much more depending on the specific task. This uniformity highlights the robustness of our method, as it is either the best-performing or very close to the best in all cases, regardless of task type or setting (unsupervised or semi-supervised). In the following, we analyze in detail the results of the different tasks of the Bio-ML track. 

The OMIM-ORDO task exhibits the lowest F-Measure across all tasks and approaches, highlighting the significant challenges faced in this domain. Specifically, linking disease subtype concepts is particularly difficult in the biomedical field, especially when dealing with rare diseases \citep{wang2023exploring}. A key factor contributing to this performance gap may be the limited structural information embedded in the ontologies used for the task \citep{he2022machine}. This factor may play a role in making knowledge graph embedding techniques, such as AMD \citep{wang2022amd}, less effective at predicting mappings. Another factor contributing to the difficulty of this task may be the mismatch between the entities' names and the standard biomedical nomenclature \citep{khatwani2024makes}. Specifically, the entity names in these ontologies tend to be excessively long, deviate from common syntactic structures used in medical terminology, and are infrequent in the relevant literature, as they focus on rare diseases. As a result, traditional matching methods that apply efficient string matching techniques, such as LogMapLt \citep{jimenez2017logmap}, are less effective. In contrast, approaches that leverage domain-specific knowledge, such as LogMapBio \citep{jimenez2017logmap} or BioGITOM, perform better for this task. Our findings suggest that pre-trained language models, such as SBERT \citep{reimers2019sentence}, can efficiently interpret these unconventional terminologies without the need for additional training or domain knowledge. Despite many current OM systems using the same retriever model (SBERT) as MILA, their performance does not match the results achieved by our approach. The key challenge lies in distinguishing between SBERT candidates that are semantically similar and those that merely overlap statistically \citep{kolyvakis2018biomedical}. By combining RAG techniques with advanced search strategies, MILA provides a more effective solution to this problem.

Similarly, the SNOMED-NCIT (Neoplasm) task involves ontologies that, while containing limited structural information \citep{he2022machine}, are more aligned with established standards. As a result, the overall performance of the OM systems tends to improve. On the other hand, the NCIT-DOID task achieves the highest F-Measure score and is therefore the least challenging task \cite{he2022machine}. The adoption of a standard biomedical terminology and rich structural information within the involved ontologies provides a favorable environment for all OM systems. Despite this, MILA still outperforms current methods, achieving the best results in both unsupervised and semi-supervised environments. Although LogMapBio \citep{jimenez2017logmap} and SORBETMatcher \citep{gosselin2023sorbet} achieve similar results, MILA has several advantages. Specifically, LogMapBio makes use of knowledge specific to the biomedical domain, whereas MILA is domain-agnostic, so it could be applied to align ontologies from other domains. Moreover, although SORBETMatcher is also supported by SBERT, it requires a fine-tuning stage, whereas MILA is a zero-shot approach that does not require any training data. 

The SNOMED-FMA and SNOMED-NCIT (Pharm) tasks require the recognition of lexical patterns within the target ontologies. For example, patterns such as \textit{set of $<$something$>$} are usual in FMA or \textit{$<$something$>$-containing product} in NCIT. As a result, these tasks are well suited to automated learning-based methods. For example, Matcha-DL leverages a linear neural network to effectively classify candidate mappings \citep{cotovio2024matcha}. However,  when MILA was applied to these tasks, we observed a higher proportion of ambiguous mappings compared to those in more successful tasks. This increased ambiguity led to a higher number of edge cases and then a greater dependence on LLM responses, which were less effective. Despite this, MILA's performance is in line with that of supervised approaches, highlighting its potential even in scenarios more suited to supervised methods.

\subsection{Limitations and Future Work}

Although MILA present improvements in performance, there are still areas for future work. Currently, MILA involves the use of a greedy search strategy based exclusively on prediction values, which may not achieve the best solution. Future work will explore other informed search strategies that combine these values with structural information from ontologies \citep{jimenez2013evaluating}. Moreover, the use of simple prompts limits the full potential of LLMs. We plan to conduct further experiments that test innovative techniques for generating prompts that incorporate contextual ontology information \citep{sampels2024exploring} or relevant examples \citep{hertling2023olala}. Furthermore, although MILA significantly reduces the number of LLM queries, compared to existing LLM-based OM systems, it could still benefit from further optimization in terms of parallelization by integrating the framework proposed by \citep{jimenez2020dividing} into our pipeline, especially for very large ontologies. In addition, our current experiments focus on simple pairwise OM \citep{euzenat2007ontology}. However, more sophisticated tasks, such as predicting subsumption relations \citep{he2023deeponto} or complex mappings \citep{silva2024complex}, present additional challenges. In the near future, we intend to explore how these tasks can benefit from the integration of LLMs and state-space search algorithms.

Finally, a key challenge with embedding-based methods is their scalability. This challenge goes beyond LLM-based OM systems, as currently most OM systems are using or combining these methods \citep{wang2022amd, cotovio2024matcha, gosselin2023sorbet}. As we have seen in our study, the process of generating embeddings and performing similarity searches can become computationally expensive, particularly for large ontologies with numerous entities. As the size of the ontologies grows, both the generation of embeddings and, more critically, the search for corresponding entities become progressively resource-intensive. This growing demand for computational resources can restrict the effectiveness and applicability of embedding-based methods in large-scale OM tasks, where high efficiency and scalability are needed. To improve scalability for large ontologies, MILA will test several optimizations. A recent solution proposes replacing the brute-force search with an approximate k-nearest neighbor search, using cosine similarity as the distance metric \citep{HybridOM}. This option significantly reduces the retrieval time complexity, although at the cost of losing some accuracy. We will also test other optimizations that include the combination of logic-based modules to manage large ontologies more efficiently \citep{jimenez2020dividing}.
\section{Conclusion} 
In summary, MILA represents a significant step forward in the development of scalable, high-performance LLM-based OM systems. By combining RAG and search strategies, MILA offers an effective solution to the challenges of LLM-based OM systems by improving computational overhead and performance in critical domains, such as the biomedical domain. MILA also offers an effective solution to the challenges of OM in general, by exhibiting task-agnostic performance without the need for training data, making it a promising approach for LLM-based OM applications. Future research could focus on further enhancing the scalability of MILA and expanding its applicability
to a wider range of domains and tasks.

\section{CRediT authorship contribution statement}
Maria Taboada$\:$ Writing – original draft, Implementation, Validation, Methodology. Diego Martinez: Writing – review $\&$ editing, Methodology. Mohammed Arideh: Implementation, Validation. Rosa Mosquera: Writing – review $\&$ editing, Funding acquisition.

\section{Declaration of competing interest}
The authors declare that they have no known competing financial interests or personal relationships that could have appeared to
influence the work reported in this article.

\section{Acknowledgments}
The authors appreciate the support and training from the University of Santiago de Compostela, as well as the support from the projects AF4EU (101086563) and SUS-SOIL (101157560), funded by the European Union's Horizon Europe program. The authors also express their gratitude to Diego Martinez-Taboada for insightful conversations.

\section{Data availability}
\texttt{https://github.com/mariatab/MILA}

\appendix

\section{State-of-the-Art OM Systems}
\label{app3}
Table \ref{tab:Approaches} provides a detailed overview of the state-of-the-art OM systems used for comparison in the MILA evaluation. The selected systems span a range of methodologies, including machine learning-based approaches, RAG frameworks, and traditional heuristic methods. Specifically, Table \ref{tab:Approaches} focuses on the following characteristics:

\begin{itemize}
    \item The \textit{type of ontology knowledge} used to predict the mapping candidates. Examples are textual knowledge and structure knowledge. Textual knowledge can include preferred names, synonyms, annotations, etc. Structure knowledge can include either any type of relationship defined in the ontology or only hierarchical relationships.
    \item The \textit{prediction model} used to generate the mapping candidates. Most OM systems include pre-trained learning models and string-based matchers.
    \item The inclusion of a \textit{fine-tuning} stage to adapt the pre-trained learning model to mapping prediction. 
    \item The type of mapping refinement. Examples include heuristic-based refinement, logical reasoning-based refinement and LLM-based refinement. 
\end{itemize}

\begin{table}[ht]
\begin{center}
\begin{tabular}{| c | c | c | c | c | c |}
    \hline
    OM  & Ontology& Prediction &Fine- &Domain-& Mapping \\
    System &  Knowledge&  Model&Tuning &specific& Refinement \\
    & in Prediction & & & Knowledge& \\\hline
    AMD & Textual and&	SBERT &  Yes&  No& Heuristic\\
    &structural &and TransR && &filtering\\
    &knowledge& && &\\
    BERTMap & Textual&Lexical indexation,&Yes &No&Logical reasoning \\
    &knowledge& String-based matching& &&based extension \\
    &&and BERT& &&and filtering\\
    BERTMapLt &Textual&Lexical indexation& No&No& No\\
    &knowledge&String-based matching& &&\\
    &&and BERT& &&\\
    BioGITOM & Textual and & BioBERT and & No & Yes & No\\
    &structural &GNNs && &\\
    &knowledge& && &\\
    BioSTransMatch & Textual & BioClinicalBERT & Yes &Yes & Heuristic\\
    &knowledge& & &&filtering\\
    HybridOM & Textual and & gtr-t5-large, k-NN  & No &No& No \\
    &hierarchical& search, BM25& &&  \\
    &knowledge& & &&  \\
    LLMs4OM  &Textual and & OpenAI & No&No&LLM-based\\ 
    &hierarchical& text-& &&and heuristic\\
    &knowledge& embedding-ada& &&filtering\\
    LogMap & Textual and &Lexical indexation,&No&No &Logical reasoning \\
    &structural& String- and& &&based extension \\
    &knowledge&Structure-based matching& &&and repair\\
    LogMapBio & Textual and &Lexical indexation,&No &Yes&Logical reasoning \\
    &structural& String- and&& &based extension \\
    &knowledge& Structure-based matching&& &and repair\\
    LogMapLt  & Textual and &Lexical indexation,&No &No&No \\
    &structural & string-based matching& &&\\
    &knowledge&& &&\\
    Matcha & Textual&String-based  matchers&No&No &Heuristic\\
    &knowledge& and SBERT& &&filtering\\
    Matcha-DL & Textual &String-based matchers&Yes &No&Heuristic\\
    &knowledge& and SBERT& &&filtering\\
    OLaLa & Textual & String-based matcher & No &No&LLM-based and\\
    &knowledge& and SBERT& &&heuristic filtering\\
    SORBETMatcher  &Textual and& SBERT  & Yes&No& Heuristic \\
    &hierarchical && &&filtering\\
    & knowledge&& &&\\
    MILA & Textual & SBERT & No&No&PDFS with \\
    &knowledge&& &&retrieve-identify-\\
    &&& &&prompt pipeline\\\hline
\end{tabular}
\caption{Overview of all OM systems that have evaluated the complete biomedical dataset}
\label{tab:Approaches}
\end{center}
\end{table}
\bibliographystyle{elsarticle-num} 
\bibliography{cas-refs}



\end{document}